\documentclass[12pt]{article}
\usepackage{amsmath,amssymb,graphicx}
\title{Traversable wormholes in a string cloud}
\author{Mart\'{\i}n G. Richarte and Claudio Simeone\thanks{E-mail: csimeone@df.uba.ar. }\\
{\small  Departamento de F\'{\i}sica, Facultad de Ciencias Exactas y 
Naturales, UBA}
   \\ 
{\small Ciudad Universitaria, Pab. I, 1428, 
Buenos Aires, Argentina}}

\begin{document}

\maketitle

\begin{abstract}

\noindent We study spherically symmetric thin-shell wormholes in a string  cloud background  in (3+1)-dimensional spacetime.
The  amount  of exotic matter required for the construction, the traversability  and the stability  under radial perturbations, are  analyzed as  functions of the parameters of the model. Besides, in the Appendices a non perturbative approach to the dynamics and a possible extension of the analysis to a related model are briefly discussed.

\end{abstract}

\baselineskip.25in
\section{Introduction}

 After the well known leading paper by Morris and Thorne \cite{motho}, considerable attention has been devoted to the study of  traversable Lorentzian wormholes  \cite{visser}.  Such kind of geometries would  connect two regions of the same universe, or of two universes, by a traversable  throat.  If they actually exist, they would present some features of particular interest as, for example, the possibility of using them for  time travel \cite{morris,novikov}. However, the  flare-out condition \cite{hovis1} to be  satisfied at the throat  requires the presence of exotic matter, that is, matter which violates the null energy condition (NEC) \cite{motho,visser,hovis1,hovis2}. However, it was shown in Ref. \cite{viskardad}   that the amount of exotic matter necessary for the existence of a wormhole can be made infinitesimally small by  suitably choosing  the geometry. After this,  special attention has been devoted to quantifying the amount of exotic matter  \cite{bavis,nandi1}; in particular, this  amount  has been pointed as an indicator of the physical viability of a traversable wormhole \cite{nandi2}. 
Besides, for a wormhole to be considered traversable, the geometry must be such that the magnitude of tidal forces are admissible for an hypothetic traveller; this has also been analyzed for most physically meaningful configurations studied in the literature.  

Of course, no unstable solution of the equations of gravitation could be of interest as a candidate for a traversable wormhole. Thus, besides the characterization of static solutions, their stability under perturbations must always be explored. In particular, this has been thoroughly studied for the case of small perturbations preserving the symmetry of the original configuration. A class of wormholes for which Poisson and Visser \cite{poisson} developed a straightforward approach for analyzing this aspect  are thin-shell ones, that is, wormholes which are mathematically constructed by cutting and pasting two manifolds to obtain a geodesically complete new manifold \cite{mvis}. In these wormhole configurations  the exotic matter lies in a shell placed at the joining surface; so the theoretical framework for dealing with them is the Darmois--Israel formalism, which leads to the Lanczos equations, that is, the  Einstein's equations  projected on the joining surface \cite{daris,mus}. Once an equation of state for the exotic matter in the shell is provided, the solution of the Lanczos equations gives the dynamical evolution of the wormhole.  Such a procedure has been subsequently followed  to study the stability of more general spherically  symmetric configurations (see, for example, Refs. \cite{eirom}).

According to present day theoretical developments, a scenario in which the fundamental building blocks of nature are extended objects instead of point objects should be considered quite seriously. In particular, 1--dimensional objects (strings)  are the most popular candidate for such fundamental objects. The study of the gravitational effects of matter in the form of clouds of both cosmic and fundamental strings has then deserved considerable attention; see for example Refs. \cite{varios}. We are interested in the viability of wormholes; so within this framework any reasonable configuration including more parameters and thus allowing for improving its features as: amount of exotic matter, pressure, traversability and stability, deserves to be analyzed.   In the present work we  start from the metric proposed in the  leading paper by Letelier  \cite{letelier} to address the study of thin-shell wormholes associated to a string cloud (and also  a global monopole; see below). We  study in detail the amount of exotic matter required for the construction, the traversability and also the stability of the configuration under perturbations preserving the original symmetry. As we shall see, comparing with the  Schwarzschild case, the string cloud allows for more freedom in the choice of the configurations to be stable, and also allows to reduce the amount of exotic matter without increasing the pressure. Because recently certain attention was devoted to exotic matter fulfilling the Chaplygin gas equation of state, in  Appendix A we explicitly impose it on the shell matter to obtain its time evolution beyond a perturbative approach. In Appendix B we discuss a possible extension of the analysis to a related geometry.  Throughout the paper we set units so that $c=G=1$.

\section{Wormholes in a string cloud with spherical symmetry}

\subsection{The string cloud}

The action of a string evolving in spacetime is given by 
\begin{equation}
S=\int{\cal L}\,d\lambda^0\,d\lambda^1,\ \ \ \ \ \ \ {\cal L}=m\sqrt{-\gamma},
\end{equation}
where $m$ is a constant characterizing each string, $\lambda^0$, $\lambda^1$ are a timelike and a spacelike parameter, and $\gamma$ is the determinant of the induced metric on the string world sheet:
\begin{equation}
\gamma=\det \gamma_{AB},\ \ \ \ \ \ \ \gamma_{AB}=g_{\mu\nu}(x)\frac{\partial x^\mu}{\partial\lambda^A}\frac{\partial x^\nu}{\partial\lambda^B}.\end{equation}
Introducing the bivector 
\begin{equation}
\Sigma^{\mu\nu}=\epsilon^{AB}\frac{\partial x^\mu}{\partial\lambda^A}\frac{\partial x^\nu}{\partial\lambda^B}
\end{equation}
with $\epsilon^{AB}$ the two-dimensional Levi--Civita symbol, the Lagrangian density ${\cal L}$ can be put as
\begin{equation}
{\cal L}=m\left(-\frac{1}{2}\Sigma^{\alpha\beta}\Sigma_{\alpha\beta}\right)^{1/2}.
\end{equation}
With this notation, a cloud of strings is described by the energy-momentum tensor
 \begin{equation}
T^{\mu\nu}=\rho_0\Sigma^{\mu\beta}\Sigma^\nu_\beta(-\gamma)^{-1/2},
\end{equation}
where $\rho_0$ is the proper density of the cloud. The quantity  $\rho_0(-\gamma)^{1/2}$ is gauge invariant, and is called the {\it gauge invariant density} of the cloud \cite{letelier}. In the case of a static spherically symmetric cloud, we have
\begin{equation}
\rho_0(-\gamma)^{1/2}=\frac{a}{r^2}\label{rho0}
\end{equation}
with $a$ a positive constant.

The general solution to Einstein's equations for a string cloud with spherical symmetry in (3+1)-dimensional spacetime, that is with density given by (\ref{rho0}), takes the form \cite{letelier}
\begin{equation}
{ds}^{2}=-f(r)\,dt^{2}+\frac{1}{f(r)}\,dr^{2}+r^{2}(d\theta^2+\sin^2\theta d\varphi^2) \label{1}
\end{equation}
where 
\begin{equation}
f(r)=1-a-\frac{2M}{r}.
\end{equation}
This metric represents  the spacetime associated with a spherical  mass $M$ centered at the origin of the system of coordinates, surrounded by a spherical cloud of strings. Besides, it can also be understood as the metric associated to a global monopole, which increases the interest of starting from it to construct a wormhole (for the details see Refs. \cite{barriola,dando}, and for a related work involving a traversable wormhole see Ref. \cite{ghosh}).  The event horizon of this metric is placed at
\begin{equation}
r_{hor}=\frac{2M}{1-a} 
\end{equation}
where $a\neq 1$. If $a$ is less than unity we have that the cloud of strings enlarges the Schwarzschild radius of the mass by the factor $1/(1-a)$.
When $a>1$ the metric represents a homogenous spacetime. The cloud of strings alone $(M=0)$ does not have horizon; it only presents a naked singularity at $r=0$.
In Fig. 1 we show where the  event horizon is located  and how this  changes with the parameter $a$. When $a=0$ we recover the Schwarzschild radius  and for $a$  close to unity the event horizon radius  tends to infinity.

\begin{figure}[h]
\begin{center}
\includegraphics[height=10cm]{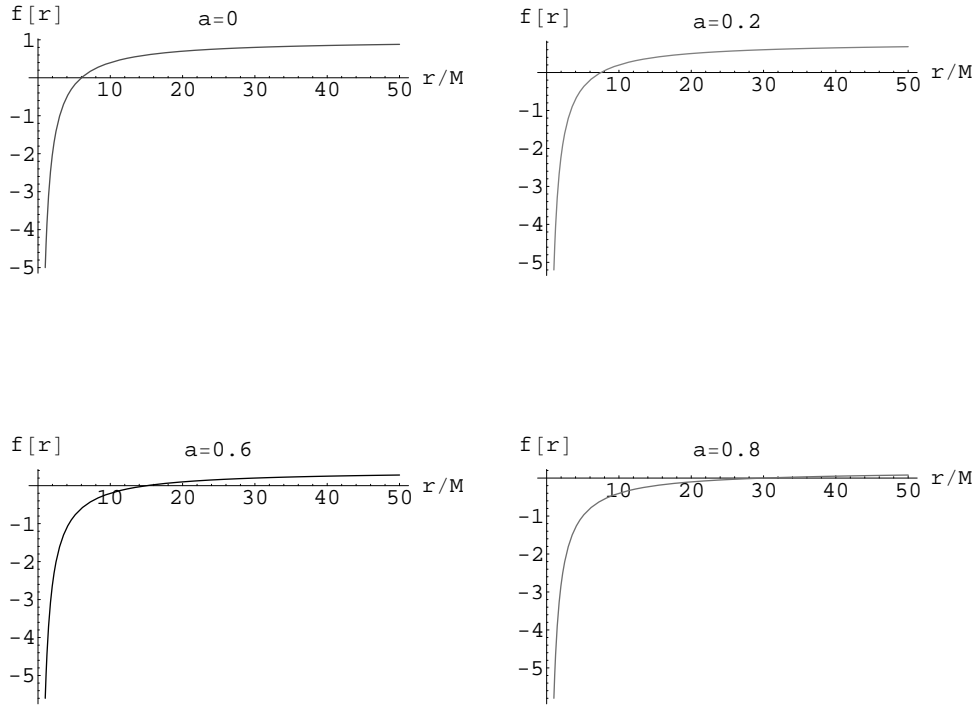}
\caption{The function $f(r)$ is shown. We can see how the position of the event horizon changes when the parameter $a$ varies within the range $[0,1)$}
\end{center}
\end{figure}

\subsection{Wormhole construction}

Now we build a spherically symmetric thin-shell wormhole starting from the generic geometry (\ref{1}) (see Fig. 2). 
We take two copies of the string cloud spacetime and remove from each manifold the four-dimensional regions described by
\begin{equation}
{\cal M}_{1,2}=\left\{r_{1,2}\leq{b}| b>r_{hor}\right\}.
\end{equation}
 The resulting manifolds have boundaries given by  the timelike hypersurfaces
\begin{equation}
\Sigma\equiv\Sigma_{1,2}=\left\{r_{1,2}={b}|b>r_{hor}\right\}.
\end{equation}
Then we paste or identify these two timelike hypersurfaces to obtain a geodesically complete new manifold $\cal M$ with a matter shell  at the surface $r=b$, where the throat of the wormhole is located. This manifold is constituted by  two asymptotically {\it locally} flat\footnote{The spacetime presents a deficit solid angle; see Ref. \cite {dando}.}  regions connected by a traversable Lorentzian wormhole.
\begin{figure}[h]
\begin{center}
\includegraphics[height=8cm]{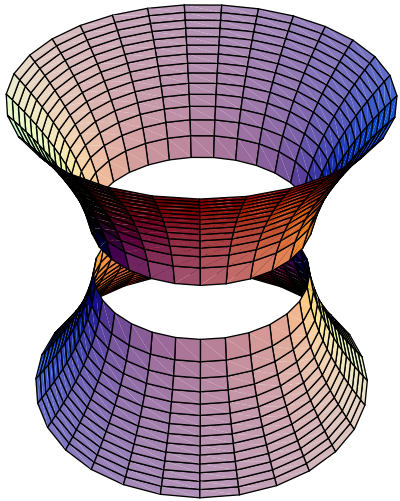}
\caption{We show two copies of  the geometry represented by Eq. (1), with $\theta=\pi/2$ and $t=const.$, for $r>r_{hor}$ when $a=0$. After the boundaries are identified we get a geodesically complete new manifold with a matter shell at $r=b$.}
\end{center}
\end{figure}
\vspace{0.5cm}

To study this class  of  wormhole we use the standard Darmois--Israel formalism \cite{daris,mus}.
The wormhole throat is placed at the surface  $\Sigma$. This  is  a synchronous timelike hypersurface. We can introduce coordinates $\xi^{i}=(\tau,\theta,\varphi)$ in $\Sigma$, with $\tau$ the proper time on the throat. In order to be able to perform  a quite general  analysis (for example, to study  the mechanical stability of the configuration; see below), we allow the radius of the throat  be a function of the proper time, $b=b(\tau)$. This is strictly right because the existence of a generalized Birkhoff theorem for the string cloud \cite{letelier} ensures that the  metric of the embedding remains the same independently of the motion of the throat, as long as the spherical symmetry is preserved.  So, the boundary hypersurface reads: 
\begin{equation} 
\Sigma: {\cal F}(r,\tau)=r-b(\tau)=0.
\end{equation} 
The second fundamental form (extrinsic curvature) at the two sides of the throat is given by
\begin{equation}
{K}^{\pm}_{il}=-n^{\pm}_{\gamma}\left(\frac{\partial^{2}X^{\gamma}}{\partial\xi^{i}\partial\xi^{l}}+\Gamma^{\gamma}_{\alpha\beta}\frac{\partial X^{\alpha}}{\partial\xi^{i}}\frac{\partial X^{\beta}}{\partial\xi^{l}}\right)
\end{equation} 
where $n^{\pm}_{\gamma}$ are the unit normals ($n_{\gamma}n^{\gamma}=1$) to the surface $\Sigma$ in $\cal M$. Defining the jump in the extrinsic curvature as $[{K}_{il}]={K}^{+}_{il}-{K}^{-}_{il}$ and its trace as
${K}=Tr([{K}_{il}])$ we obtain the so-called Lanczos equations:
\begin{equation}
-[{K}_{il}]+{K}g_{il}=8\pi{S_{il}}
\end{equation}
where $S_{il}$ is the surface stress-energy tensor of the shell  placed at the throat.
The non vanishing components  of the extrinsic curvature are:
\begin{equation}
{K}_{\tau\tau}^{\pm}=\pm
\frac{\ddot{b}+f^{'}(b)/2}{\sqrt{f(b)+\dot{b}^{2}}},
\end{equation} 
\begin{equation}
{K}_{\theta\theta}^{\pm}={K}_{\varphi\varphi}^{\pm}=\pm\frac{2}{b}{\sqrt{f(b)+\dot{b}^{2}}}, 
\end{equation} 
where the dot means derivation with respect to the proper time $\tau$, and a prime stands for a derivative with respect to $r$. Then, from Eqs. (8), (9) and (10) we get a formal expression for the pressure $p=S_\theta^\theta=S_\varphi^\varphi$ and the energy density $\sigma=-S_\tau^\tau$ in terms of $b(\tau)$, first and second derivatives of $b(\tau)$, and the function $f$ which depends on the parameters of the system $(a, M)$:
\begin{equation}
\sigma=-\frac{1}{2\pi b}\sqrt{f(b)+\dot{b}^{2}},
\end{equation}
\begin{equation}
p=-\frac{1}{2}\sigma + \frac{1}{8\pi}\frac{\ddot{b}+f^{'}(b)/2}{\sqrt{f(b)+\dot{b}^{2}}},
\end{equation}
where the prime means derivation with respect to $r$.
As it was to be expected, the energy density is negative, indicating the existence of exotic matter at the shell. We shall focus on this aspect of the wormhole in the next Section.
It is easy  to see from Eqs. (11) and (12) that the energy conservation equation is fulfilled:
\begin{equation}
\frac{d(A\sigma)}{d\tau}+p\frac{d A}{d\tau}=0,
\end{equation}
where $A$ is the area of the wormhole throat. The first term in Eq. (13) represents the internal energy change of the shell and the second the work by internal forces of the shell.
The dynamical evolution of the wormhole throat is governed by the Lanczos equations and to close the system  we must supply an equation of state $p=p(\sigma)$ that relates $p$ and $\sigma$.  In Section 4 we shall develop this point in a perturbative approach, while in  the Appendix we shall obtain  the dynamical evolution of the throat for a particular equation of state.

\section{Characterization of the construction}

\subsection{Amount of  exotic matter}

Many authors (see for instance Refs. \cite{nandi1,nandi2,loboe,eisi05}) have proposed to quantify the amount of exotic matter  as a way to characterize the viability  of  a  traversable wormhole. Here we shall analyze the energy conditions and evaluate the total amount  of exotic matter for the wormholes built in Section 2, in the case of static configurations, i.e., $b=b_{0}$. In this case, the energy density and pressure are
\begin{equation}
\sigma_0=-\frac{1}{2\pi b_0}\sqrt{1-a-\frac{2M}{b_0}},
\end{equation}
\begin{equation}
p_0=\frac{1}{4\pi b_0}\sqrt{1-a-\frac{2M}{b_0}}+\frac{M}{8\pi b_0^2}\frac{1}{\sqrt{1-a-\frac{2M}{b_0}}}.
\end{equation}
The {\it weak energy condition} (WEC) states that for any timelike  vector $U^{\mu}$ it must be $T_{\mu\nu}U^{\mu}U^{\nu}\geq0$; the WEC also implies, by continuity, the {\it null energy condition} (NEC), which means that for any null vector $k^{\mu}$ it mus be $T_{\mu\nu}k^{\mu}k^{\nu}\geq0$
\cite{visser}. In an orthonormal basis the WEC reads $\rho\geq0$, $\rho+p_{l}\geq0\  \forall\, l$, while the NEC takes the form $\rho+p_{l}\geq0 \ \forall\, l$. In the case of the  wormhole constructed in Section 2  we have that the radial pressure is zero, $p_{r}=0$, and the energy density verifies $\sigma<0$, so that both energy conditions are violated. The transverse pressure is $p_{t}=p$ and the sign of $\sigma+p_{t}$ is not fixed by this conditions, but it depends on the values of the parameters of the system.

There have been several proposals for quantifying the amount of exotic matter in  wormholes. In order to allow for an immediate comparison with the results of other works, we shall adopt the most usual choice, which is the  integral over space including the pressure associated to the violation of the energy conditions:
\begin{equation}
\Omega= \int (\rho + p_{r})\sqrt{-g}\,d^{3}x,
\end{equation}
where $g$ is the determinant of the metric tensor. The advantages of this quantifier,  compared with others including other measures as $\int (\rho + p_{r})\,d^{3}x$ or $\int (\rho + p_{r})\sqrt{-\,^3\hskip-.05cm g}\,d^{3}x$, have been carefully discussed in \cite{nandi1}. Besides, this choice is consistent with previous proposals for covariant conservation laws in General Relativity (see \cite{tolman-komar}). 

We introduce a new radial coordinate ${\cal R}=\pm(r-b_{0})$ with $\pm$ corresponding to each side of the shell. 
In our construction the shell does not exert radial pressure, and the energy density is located on the surface so the energy density  can then be written as $\rho=\delta({\cal R})\,\sigma_{0}$.
This yields  the following formula for the amount of exotic matter:
\begin{equation}
\Omega= \int^{2\pi}_{0}\int^{\pi}_{0}\int^{+\infty}_{-\infty}\delta({\cal R})\,\sigma_{0} \sqrt{-g}\, d{\cal R}\,d\theta\, d\varphi
      = 4\pi \sigma_{0} b^2_{0}.
\end{equation}
Replacing the explicit form of $\sigma_{0}$ we obtain the exotic matter amount as a function of the parameters ($a, M, b_{0}$) that characterize the configurations:
\begin{equation}
\Omega= -2b_{0}\sqrt{1-a-\frac{2M}{b_{0}}}.
\end{equation}
The amount of exotic matter is thus always smaller than in the Schwarzschild case $a=0$. In Fig. 3 we show   ${\Omega}/{M}$ as a fuction of ${b_{0}}/{M}$ for different values of $a$. Note that  for $b_{0}\gg M$ the amount $\Omega$ becomes linear in ${b_{0}}$:  ${\Omega}\simeq -2{b_{0}}\sqrt{1-a}$. 

\begin{figure}[h]
\centering
\includegraphics[height=10cm]{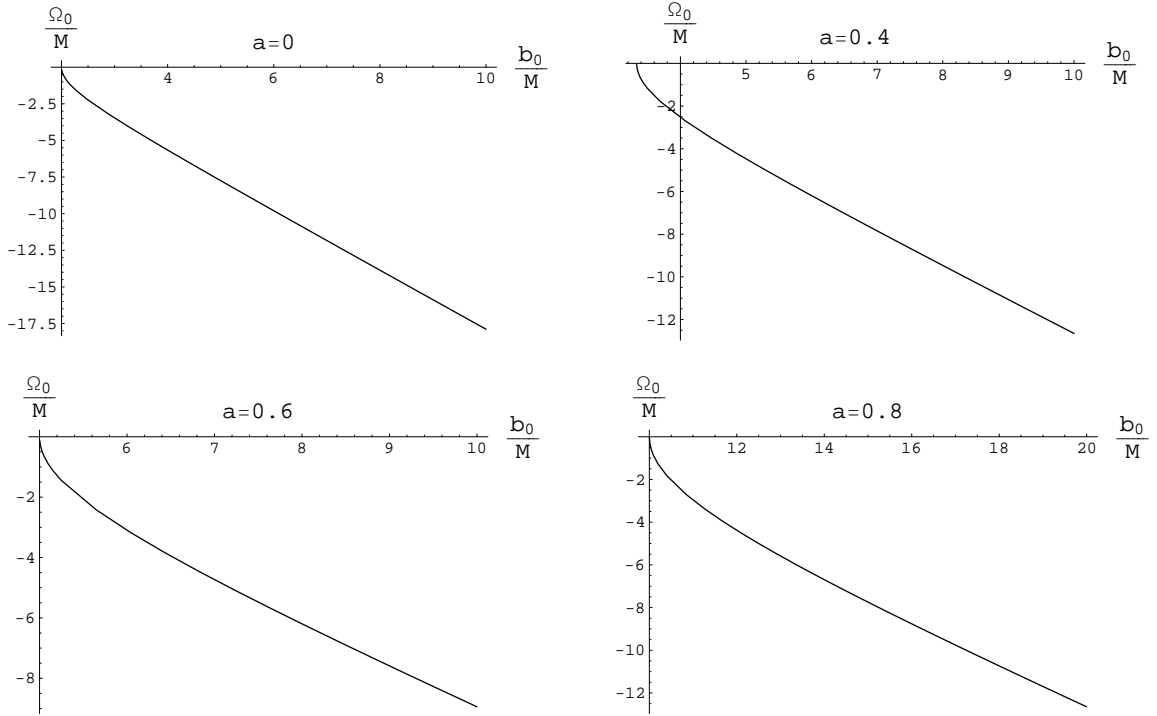}
\caption{ The exotic matter amount is shown as a function of $b_{0}$, for  given values of the parameter $a$.}
\end{figure}
\vspace{0.2cm}

A natural question is  which conditions allow to reduce  ${\Omega}$. Since ${\Omega}\propto \sigma_{0}$ and $\sigma_{0}\rightarrow0$ as $b_{0}\rightarrow r_{hor}$, we shall investige this limit more carefully. For $b_{0}$ near the event horizon $r_{hor}$, the transverse pressure and the surface density energy behave as
\begin{equation}
\sigma_{0}=-\frac{1}{2\pi r_{hor}}\sqrt{f^{'}(r_{hor})}\sqrt{b_{0}-r_{hor}}+ {\cal O}[(b_{0}-r_{hor})^{3/2}],
\end{equation}
\begin{equation}
p_{0}=\frac{1}{8\pi}\frac{\sqrt{f^{'}(r_{hor})}}{\sqrt{b_{0}-r_{hor}}}+ {\cal O}[(b_{0}-r_{hor})^{1/2}],
\end{equation} 
where $f^{'}(r_{hor})\neq0$. Then, when  we take the limit $b_{0}\rightarrow r_{hor}$  the surface energy density goes to zero but the transerve pressure tends to infinity (see Ref. \cite{eisi05} for a similar behaviour). 

We can also consider the case when the wormhole radius $b_0$ is fixed and fulfils $b_0\gg r_{hor}$, and the exotic matter amount is taken  as a function of the parameter $a\in[0,1)$. In this case  ${\Omega}$  can be reduced by increasing the value of the parameter $a$ (see Fig. 4), while the energy density and the pressure remain under control (see Figs. 5 and 6): in this limit we have
\begin{equation}
p_0\simeq\frac{1}{4\pi b_0}\left(\sqrt{1-a}+\frac{M}{2b_0\sqrt{1-a}}\right),
\end{equation}
where the second term is kept finite even if $a\to 1$ precisely because the condition  $b_0\gg r_{hor}$ implies $M/(2b_0\sqrt{1-a})<\sqrt{1-a}/4$. In fact, under these conditions the amount of exotic matter can be reduced without increasing the pressure, which constitutes a remarkable feature for a wormhole construction (in the pure Schwarzschild case and in the same limit $b_0\gg r_{hor}$, the amount $\Omega$ can only be reduced by reducing $b_0$, which leads to an increase of the pressure). This is shown in  
  Figs.  5 and 6.

\subsection{Traversability}

A possible way to define the traversability of a Lorentzian wormhole is to compare  the relative aceleration (proportional to the tidal force) between to parts of a traveller, with the magnitude of  the Earth surface gravity. We start from the expression of the  four-velocity  of  a traveller falling towards the throat straight in the radial direction (that is with zero angular momentum) 
\begin{equation}
U^{\mu}=\left(\frac{E}{f(r)},-\sqrt{E^{2}-f(r)},0,0\right),
\end{equation}
where $E$ is its energy at infinity (basically its rest mass). Then from the definition of the geodesic deviation  we calculate  the covariant relative acceleration  between two parts of the traveller separated by a distance $X^{r}$ in  the radial direction: 
\begin{equation}
\frac{DX^{r}}{Ds}=R^{r}_{trt}X^{r}U^{t}U^{t}=\frac{-2MEX^{r}}{r^3(1-a-2M/r)},
\end{equation}
where $s$ is the proper time of the  traveller.
From this expression it can easily be shown that a choice of the parameters making the wormhole stable (see the next Section) and also reducing the amount of exotic matter, can render it traversable. For example, this happens for  $a=0.6$ and $b_0=10^{4}M_{\odot}$, which for a length of order one meter give  a tidal acceleration  which is about  Earth surface gravity.
\begin{figure}[h]
\centering 
\includegraphics[height=8cm]{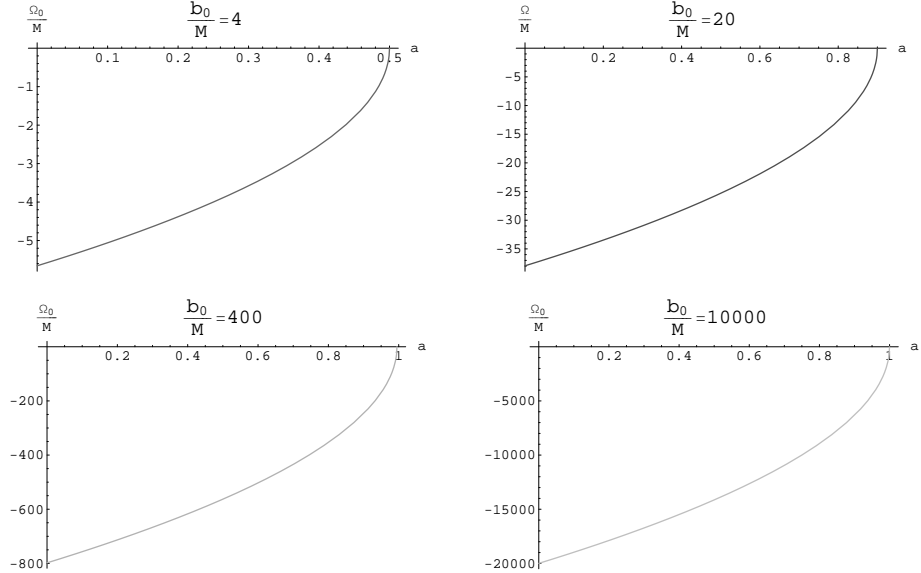}
\caption{ The exotic  matter amount is shown as a function of $a$, with $a\in [0, 0.9998)$, for  given values of $b_{0}$.}
\end{figure}

\vspace{0.5cm}
\begin{figure}[h]
\centering 
\includegraphics[height=8cm]{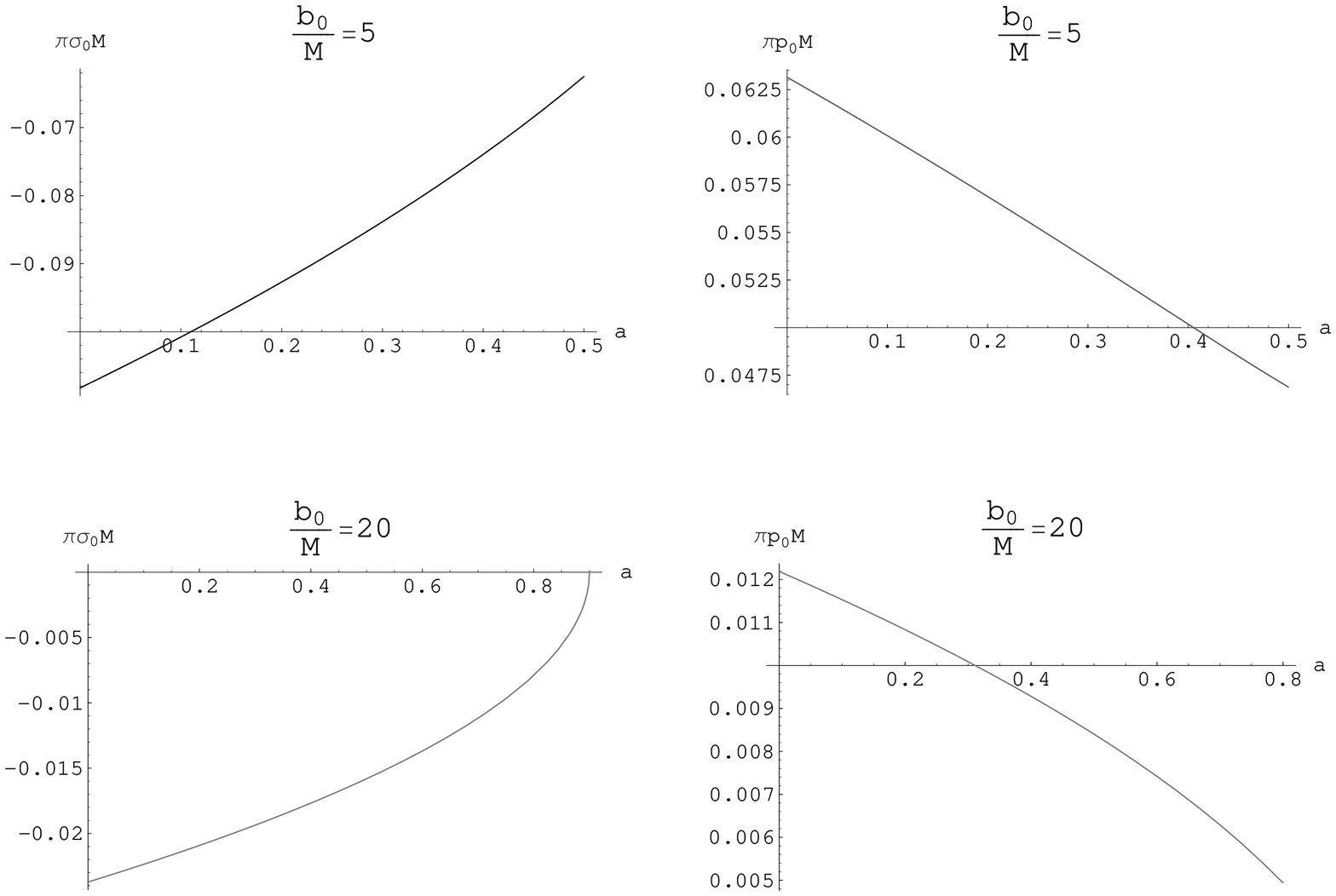}
\caption{ For  given values of the wormhole radius, the  energy density and the pressure  are plotted as functions of the parameter $a$.}
\end{figure}

\vspace{8cm}
\begin{figure}[h]
\centering
\includegraphics[height=8cm]{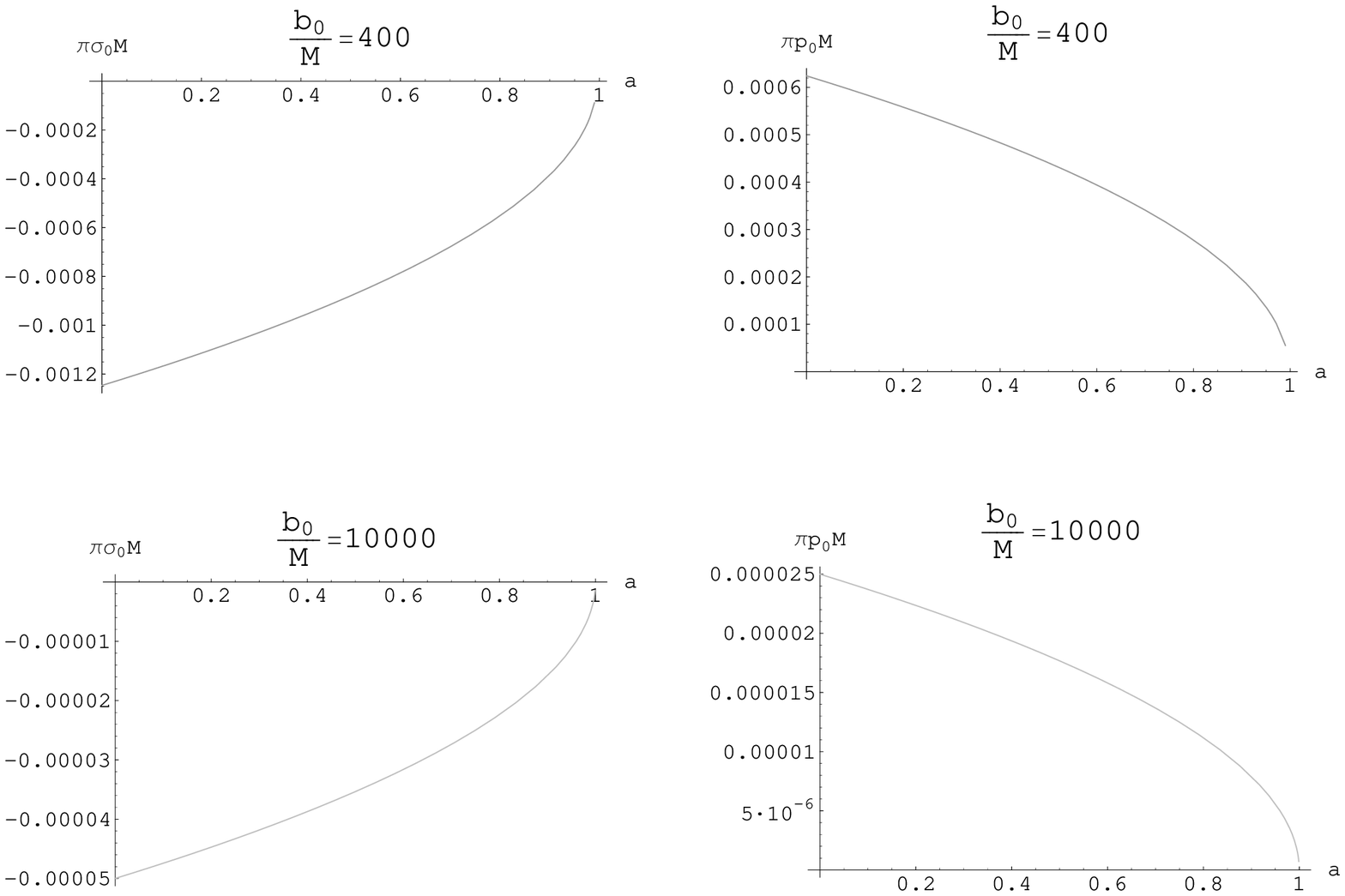}
\caption{For given values of the wormhole radius,  the density energy and the pressure  are plotted as  functions of the parameter $a\in [0,0.9998)$.}
\end{figure}

\vspace{0.5cm}

\section{Stability analysis}

A physically interesting wormhole geometry should last enough so that its traversability makes sense. Thus the stability of a given wormhole configuration becomes a central aspect of its study. Here we shall analyze the stability   under small perturbations preserving the spherical symmetry of the configuration; for this we shall proceed as Poisson and Visser in Ref. \cite{poisson}. As we said, the dynamical evolution is determined by  Eqs. (11) and (12), or by any of them and Eq. (13), and to complete the system  we must add an equation of state that relates $p$ with $\sigma $, i.e, $p=p(\sigma)$. By introducing the explicit form of the metric in Eq. (11) we have
\begin{equation}
\dot{b}^{2}-\frac{2M}{b}-[{2\pi b\sigma(b)}]^{2}=a-1.
\end{equation}
To obtain  $\sigma=\sigma(b)$ we first note that the energy conservation equation can be written  as
\begin{equation}
\dot{\sigma}=-2(\sigma+p)\frac{\dot{b}}{b}
\end{equation}
which can be integrated to give
\begin{equation}
\ln\frac{b}{b(\tau_{0})}=-\frac{1}{2}\int^{\sigma}_{\sigma({\tau_0})} \frac{d\sigma}{\sigma+p(\sigma)}\,.
\end{equation}
From Eq. (26), if the equation of state $p=p(\sigma)$ is given, one  can obtain  $\sigma=\sigma(b)$.

Following the procedure introduced by  Poisson and Visser, the analysis of the  stability of the configuration can be reduced to the analogous problem of  the stability of a particle  in  a one dimensional potential $V(b)$. This    is easy to  see if we  write  Eq. (24) as
\begin{equation}
\dot{b}^{2}= -V(b)
\end{equation}
with 
\begin{equation}
V(b)=-\frac{2M}{b}-[{2\pi b\sigma(b)}]^{2}+1-a.
\end{equation}
(We can  verify  the procedure by setting $a=0$ and see that we recover the results of Ref. \cite{poisson}).
So, to study the stability we expand up to second order  the potential $V(b)$ around the static solution $b_{0}$  (for which $\dot{b}=0,\,{\ddot{b}} = 0$). As we expect, for a stable configuration it is $V(b_{0})=0$ and $V^{'}({b_{0}})=0$, where the prime means a derivative with respect to $b$. Then, Eq. (27) takes the following form:
\begin{equation}
\dot{b}^{2}= -V^{''}(b_{0})(b-b_{0})^{2}+ {\cal O}[(b-b_{0})^{3}].
\end{equation}

To compute the  derivates it is convenient to define the parameter 
\begin{equation}
\eta(\sigma)\equiv\frac{\partial p}{\partial \sigma},
\end{equation}
 which for ordinary matter would represent the squared speed of sound: $v^2_s=\eta$. For now, however, we simply considerer $\eta$ as a useful parameter  related with  the equation of state (see below). 
Then,  we obtain the second derivative of the potencial for the  metric (1):
\begin{equation}
V^{''}(b_{0})=-\frac{2}{b^{2}_{0}f(b_{0})}\left[(2M/b_{0}) f(b_{0})+M^{2}/b^{2}_{0}+(1+2\eta_{0})f(b_{0})(1-a-3M/b_{0})\right],   
\end{equation}
where $\eta_0=\eta(\sigma_0)$.  The wormhole is stable if and only if $V^{''}(b_{0})>0$ while for $V^{''}(b_{0})<0$ a radial perturbation  grows (at least until nonlinear regime is reached) and the wormhole is unstable. Because the function $f(b_{0})$ is always positive for $b_{0}>r_{hor}$, we only have to analyze the sign of the bracket in Eq. (31) for determining which are the  values of  the parameters $M, a, b_{0}$ that make the wormhole stable. Then, after some simple manipulations, the stability conditions can be written as follows:
\begin{align}
b_{0} & >  3M/(1-a) \ \ \ \ \ \ \ \mathrm{if}   & \eta_{0}  <   -[(1-a)^{2}-3M(1-a)/b_{0} + 3M^{2}/b^{2}_{0}], \\
b_{0} & <  3M/(1-a)  \ \ \ \ \ \ \  \mathrm{if}   & \eta_{0}  >  -[(1-a)^{2}-3M(1-a)/b_{0} + 3M^{2}/b^{2}_{0}]. 
\end{align}

To get a good insight of the stability regions, we draw the curve $V^{''}(b_{0})=0$ in the plane $(\eta_{0} , b_{0}/M) $ for different values of the parameter $a$ (see Fig. 7).
We can identify the regions of stability as follows: In the case of $a=0$, for $b_{0}/M>3$ the region of stability lies under the curve, but for $b_{0}/M<3$ the stability region is placed above the curve showed in  Fig. 7 (we can also see that when $a=0$   we recover the results by Poisson and Visser). Then, for   values of $a \in [0,\,1)$ the shape of  the regions of  stability remains the same of the case   $a=0$, but their locations  are shifted as the event  horizon  changes  with the parameter $a$. We find that when $a$ is increased, the regions of stability, though shifted away, become considerably enlarged.

\begin{figure}[h]
\centering
\includegraphics[height=11cm, width=15cm]{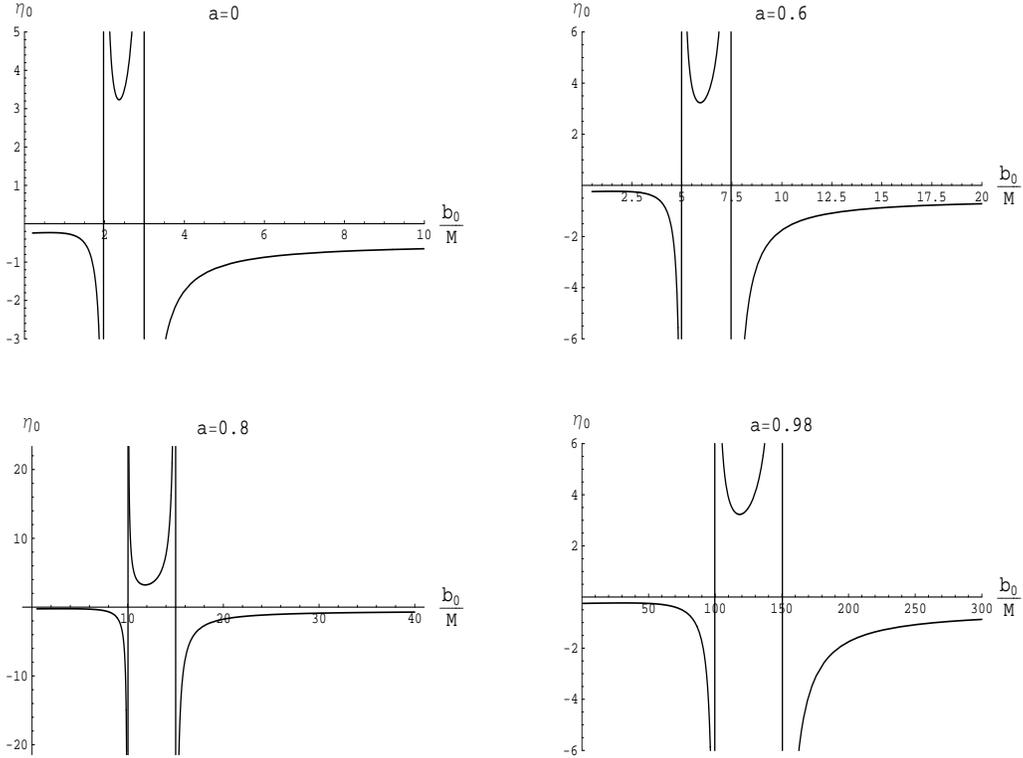}
\caption{The dependence of the stability regions  with the parameter $a$ is shown.}
\end{figure}
We observe from these plots that the stability of the wormhole configuration demands that the parameter ${\eta_{0}}$ of the exotic matter at the shell is, when positive,  greater than unity (see, however, the same analysis for a related model in Appendix B). It is clearly not easy to interprete $\sqrt{\eta_0}$ as a kind of velocity  of propagating waves at the shell.
Furthermore, from the regions  of stability obtained, we conclude that traversable wormholes  with radii $b_{0}>3M/(1-a)$ could  be  stable, under perturbations that preserve the spherical symmetry, only if the parameter $\eta_{0}$ is negative. However,  it was pointed out by Visser and Poisson that the interpretation of $\sqrt{\eta_{0}}$ as the speed of sound would require an understanding of the microphysics of  exotic matter, which is not available by now.

\section{Summary}
We have built  traversable thin-shell wormholes applying the cut and paste procedure to the  geometry corresponding to a spherical cloud of strings. We have found that the amount of exotic matter --which is restricted to the throat-- can be reduced by a suitable choice of the parameters. Moreover, we have shown that for a fixed wormhole radius   $b_0\gg r_{hor}$, there exists a range of values of the parameter $a$ such that, while the amount of exotic matter can be  reduced, the transverse pressure and the surface energy density can be kept under control. We have also studied the stability of the configuration under perturbations preserving the spherical symmetry, and we have  found that the stability regions --though shifted away--  become enlarged when the parameter $a$ is increased. Within the characterization of static configurations, we have considered the traversability  of large wormholes by evaluating the tidal force; we have found that it turns to be acceptable in the case of values of  $a$ and $b_0$ such that the wormhole is stable. Besides, in the Appendix A (see below) we briefly study the dynamics of the shell beyond the perturbative approach by considering the particular case of the Chaplygin equation of state for the shell matter. Also, the features of a related wormhole configuration are discussed in the Appendix B.

\section*{Acknowledgments}

The authors want to thank \'Alvaro Corval\'an for  useful comments on differential equations, and Ernesto Eiroa for helpful discussions.
This work was supported by  Universidad de Buenos Aires and CONICET.

\newpage

\section*{Appendix A: Dynamics of the shell}

\vspace{0.5cm}
As noted above, it  can be proved that there exists a generalized  Birkhoff theorem for the string cloud, so the geometry exhibited in Section 2  is the general solution for the case of spherical symmetry; in particular, no  gravitational waves are emitted. Thus the time evolution can be obtained from Eqs. (11) and (12) plus an equation of state relating the energy density and pressure of the exotic matter. Here, we consider the special case of the Chaplygin gas \cite{chaplin1}, i.e. a perfect fluid fulfilling
\begin{equation}
 p=-\frac{\lambda}{\sigma}
\end{equation}
where $\lambda$ is a positive constant.
A remarkable feature of the Chaplygin gas model is that it  has  positive and bounded squared sound velocity: $v^{2}_{s}=\partial p/\partial\sigma=\lambda/\sigma^{2}$, which is not trivial for an exotic  matter fluid. This model has been applied in cosmological models because it  describes a smooth transition from a deccelerated expansion of the Universe to the present epoch of cosmic acceleration and because it  gives a unified macroscopic description of dark matter and dark energy \cite{chaplin2}. It has been recently proposed for supporting a class of thin-shell wormholes in Ref. \cite{eisi07}.
\begin{figure}[hb]
\centering
\includegraphics[height=7cm, width=15.5cm]{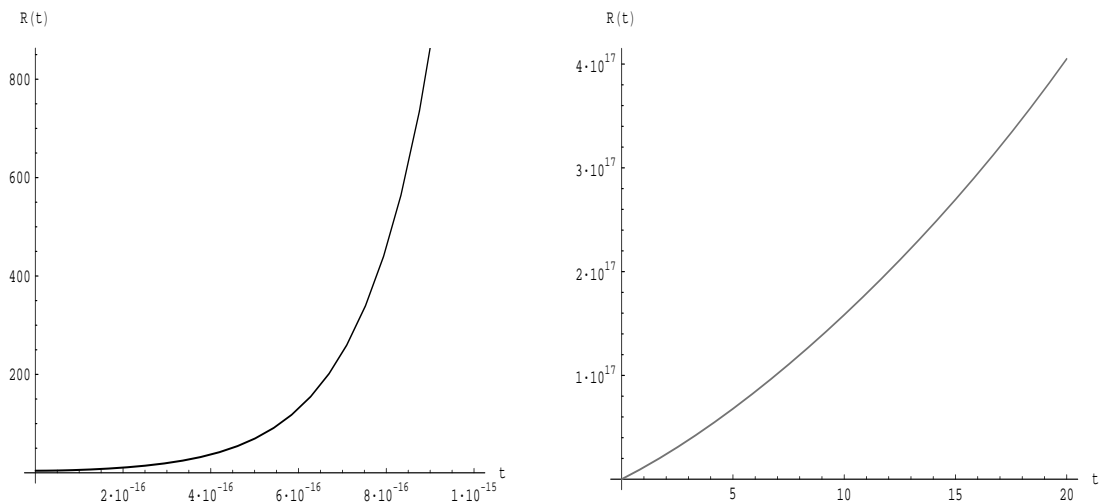}
\caption{For a given initial velocity ($\dot{R_{0}}=0$) and initial  radius ($R_{0}=3$)  we show the relation between the (scaled) proper time  $T$ and  the wormhole radius in the case of the Chaplygin equation of state. For large $R$  there is no sensible dependence with the parameter $a$ while in the case of small $R$, the evolution exhibits some dependence with $a$; here we plot $R(T)$ for $a=0.2$. Note the difference between the two plots in the scales of both axes.}
\end{figure}
If we take the Eqs. (11) and (12) and replace them in  Eq. (34), the equation for the wormhole radius $b$ can be written as:
\begin{equation}
\frac{d}{dT}\left( R\frac{dR}{dT}\right)=a-1+\frac{1}{R}+\frac{8\pi^{2}}{l^{2}_{\lambda}}R^{2},
\end{equation}
where we have defined  dimensionless variables $R={r}/M$, $T={\tau}/M$ and $\sqrt{\lambda}M=({l_{\lambda}})^{-1}$. 
In order to solve  Eq. (35), we define the variable $R=\sqrt{y}$ and  multiply this equation by $\dot{y}(T)$. So then, after integrating, we obtain a first order differential equation  for the  squared velocity:
\begin{equation}
\dot{y}^{2}(T)=\dot{y}^{2}_{0}+4(a-1)(y-y_{0})+8({\sqrt{y}}-{\sqrt{y_{0}}})+{32\pi^{2}}{l^{-2}_{\lambda}}(y^{2}-y^{2}_{0}). 
\end{equation}
Now, if we return to the  variable $R$,  Eq. (36) can be put in the following integral form:
\begin{equation}
\int^{T}_{T_{0}}dt=\pm\int^{R}_{R_{0}}\frac{RdR}{\sqrt{{(R_{0}\dot{R}_{0})^{2}}+(a-1)(R^{2}-R^{2}_{0})+2(R-R_{0})+{8\pi^{2}}{l^{-2}_{\lambda}}(R^{4}-R^{4}_{0})}}. 
\end{equation}
The solution of this integral has a closed expression in terms of elliptic functions of first and third kind. The general solution  is parametrized  by $a$, so for simplicity we just exhibit a plot of the solution for large $R$ and for small $R$ (see Fig. 8). Of course, initial conditions are taken such that $R_0$ corresponds to an initial radius greater than the horizon radius of the original manifold. In the case of large $R$, this approximation leads to study the dynamical equation  of an anti-oscillator and the solution does not exhibit a dependence with the parameter $a$. In the case of small $R$, the solution has a smooth dependence with $a$, and here we have solved the  differential equation by  Taylor expanding up to fourth order about the initial radius $R_{0}=3$. As can be seen from Fig. 8, a monotonic evolution is obtained. Thus, differing from the results obtained with a linear equation of state (Section 4), within this model no stable configurations would  exist. This seems to be consistent with the results   of \cite{eisi07}, where it was shown that in the case of the Chaplygin equation of state, stable configurations required a non vanishing charge or a cosmological constant, which, compared with the string cloud, represents a more considerable departure from the pure Schwarzschild metric.  
\newpage

\section*{Appendix B: A related model}

 An immediate extension of our analysis of Sections 3 and 4 can be performed in the case of a perfect string fluid (see \cite{letelier2} and the first paper in Ref. \cite{varios}). Such a model includes a non vanishing angular pressure such that the energy-momentum tensor of the fluid has the form $T_t^t=T_r^r=-\alpha T_\theta^\theta=-\alpha T_\varphi^\varphi$. As a result of this, for $\alpha \neq 2$ the function $f$ in the metric (1) takes the form\footnote{We choose $\alpha \neq 2$ because it allows to reproduce the results of the Schwarzchild--de Sitter and the Reissner--Nordstr\"{o}m thin-shell  wormholes in the cases $\alpha=-1$ and $\alpha=1$ respectively (see the papers by Lobo and Crawford and by Eiroa and Romero, Ref. \cite{eirom})}
\begin{equation}
f(r)=1-\frac{2M}{r}-\frac{\epsilon\alpha L^{2/\alpha}}{(\alpha -2)\,r^{2/\alpha}}
\end{equation} 
where $L$ is a positive constant of dimension length, and $\epsilon=\pm 1$ denotes the sign of the energy density of the string fluid. 
 \begin{figure}[htp]
\centering
\includegraphics[height=10cm, width=13cm]{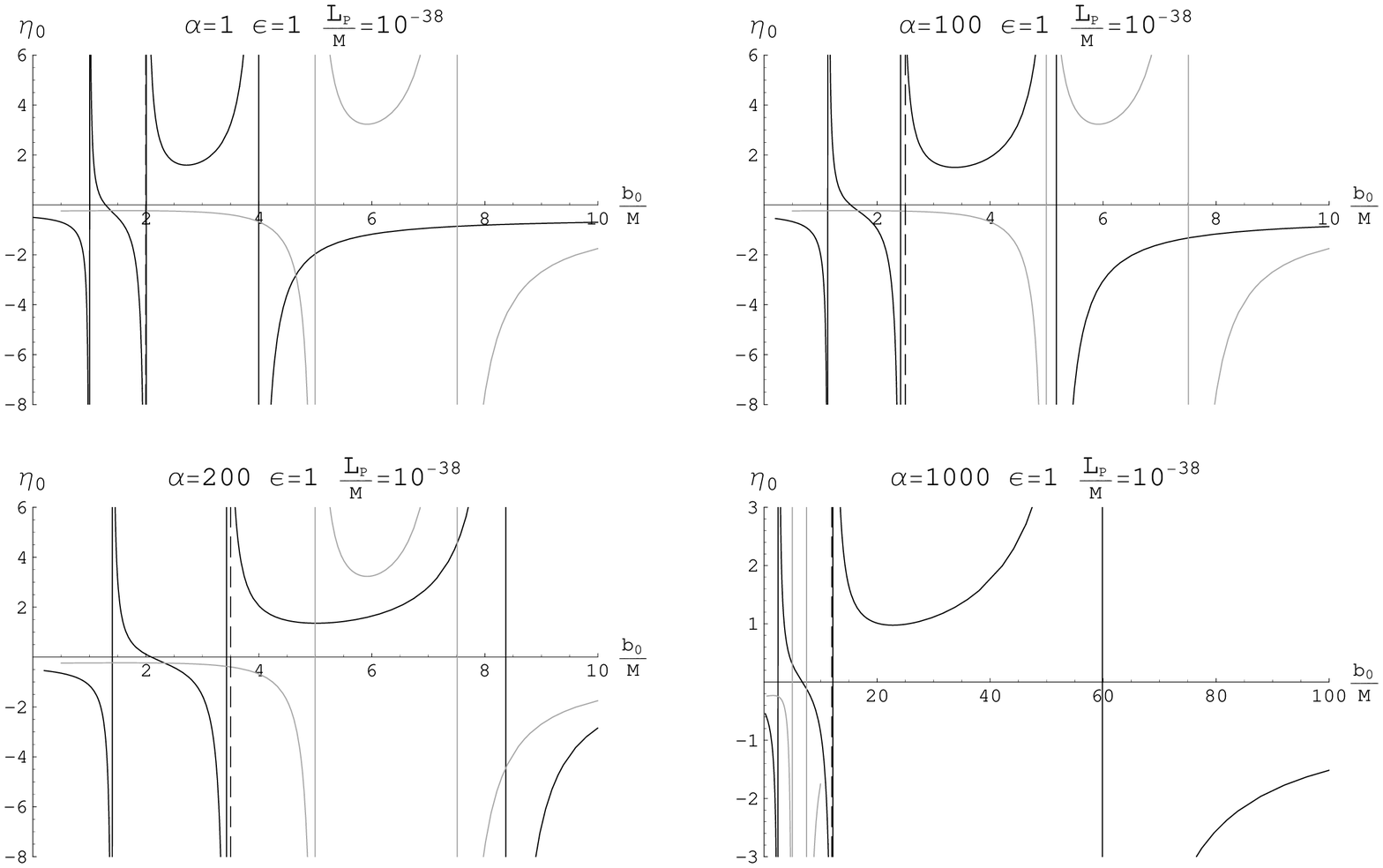}
\caption{The stability regions (black line) and the horizon (dashed line) are shown for  $\epsilon =1$, $L=10^{-38}M$ in the case of a perfect string fluid. Four values of  the  parameter $\alpha$ are considered. The stability regions are compared with those corresponding to  the cloud of strings with $a=0.6$ (gray line).}
\end{figure}
\begin{figure}[htp]
\centering
\includegraphics[height=5cm, width=13cm]{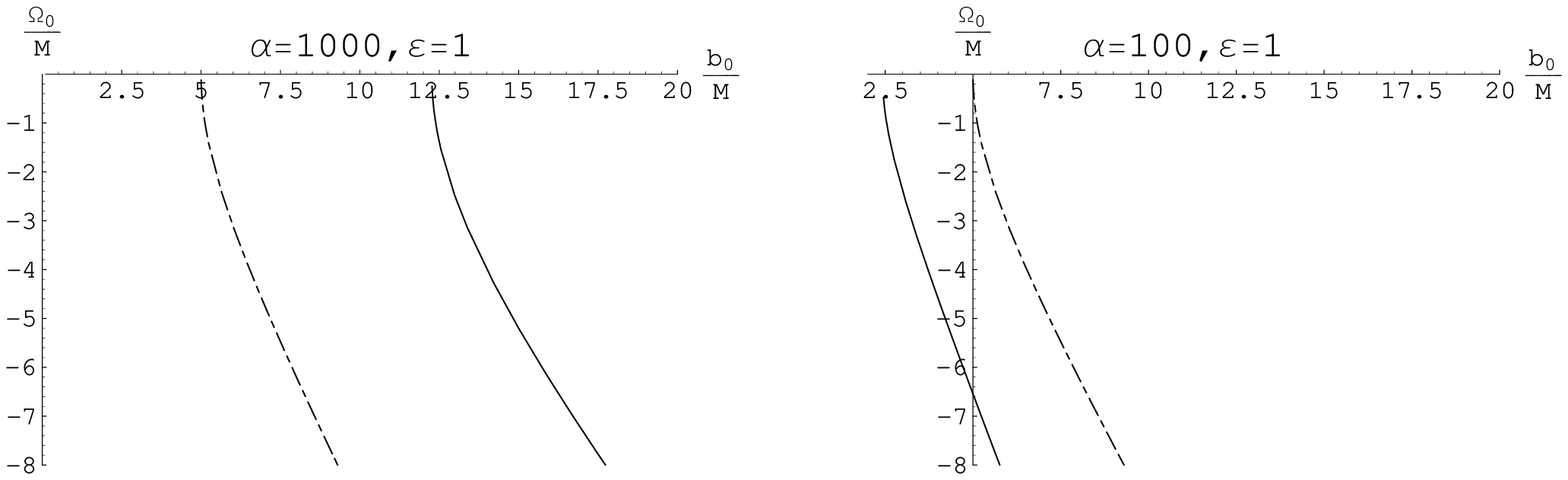}
\caption{The  amount of exotic matter is  shown for $\alpha>0$, $L=10^{-38}M$. We also show the exotic matter amount for the string  cloud (dashed line) with $a=0.6$.}
\end{figure}
Because the calculations are  analogous to those above, we omit the details and give the results for $\epsilon=+1$ (we want to restric exotic matter to the shell) and for  $L$ of order $10^{-38}M$. We compare the stability regions and the amount of exotic matter with the case of the string cloud with $a=0.6$.  As can be seen from Figs. 9 and 10, while certain values of $\alpha$ allow to reduce the amount $\Omega/M$ with respect to the string cloud case, the regions of stability for low values of $\alpha$ are shifted towards the horizon and become smaller. However,  lower values of the would be squared speed of sound become compatible with stability; in particular, for large $\alpha$ a numerical calculation shows that positive values of $\eta_0$ slightly smaller than unity are now possible.

\end{document}